# Prévision de l'épaisseur du film passif d'un acier inoxydable 316L soumis au fretting corrosion grâce au Point Defect Model, PDM.

# Predicting the steady state thickness of passive films with the Point Defect Model in fretting corrosion experiments.

**Jean GERINGER [a], Mathew TAYLOR[b] , Digby.D. Macdonald[b]**

[a]Bio-Tribocorrosion lab, STBio Department
Center for Biomedical and Healthcare Engineering EMSE, CNRS:UMR5307, LGF,
158 cours Fauriel
F-42023 Saint-Etienne, France
[2]Penn State University, MSE-CEST, 206A Steidle Building,
University Park 16802 PA USA

## Abstract

Some implants have approximately a lifetime of 15 years. The femoral stem, for example, should be made of 316L/316LN stainless steel. Fretting corrosion, friction under small displacements, should occur during human gait, due to repeated loadings and un-loadings, between stainless steel and bone for instance. Some experimental investigations of fretting corrosion have been practiced. As well known, metallic alloys and especially stainless steels are covered with a passive film that prevents from the corrosion and degradation. This passive layer of few nanometers, at ambient temperature, is the key of our civilization according to some authors. This work is dedicated to predict the passive layer thicknesses of stainless steel under fretting corrosion with a specific emphasis on the role of proteins. The model is based on the Point Defect Model (micro scale) and an update of the model on the friction process (micro-macro scale). Genetic algorithm was used for finding solution of the problem. The major results are, as expected from experimental results, albumin prevents from degradation at the lowest concentration of chlorides; an incubation time is necessary for degrading the passive film; under fretting corrosion and high concentration of chlorides the passive behavior is annihilated.

## Résumé

Les implants orthopédiques de hanche ont une durée de vie d'environ 15 ans. Par exemple, la tige fémorale d'un tel implant peut être réalisée en acier inoxydable 316L ou 316LN. Le fretting corrosion, frottement sous petits déplacements, peut se produire pendant la marche humaine en raison des chargements répétés entre le métal de la prothèse et l'os. Plusieurs investigations expérimentales du fretting corrosion ont été entreprises. Cette couche passive de quelques nanomètres, à température ambiante, est le point clef sur lequel repose le développement de notre civilisation, selon certains auteurs. Ce travail vise à prédire les épaisseurs de cette couche passive de l'acier inoxydable soumis au fretting corrosion, avec une attention spécifique sur le rôle des protéines. Le modèle utilisé est basé sur le Point Defect Model, PDM (à une échelle microscopique) et une amélioration de ce modèle en prenant en compte le processus de frottement sous petits débattements. L'algorithme génétique a été utilisé pour optimiser la convergence du problème. Les résultats les plus importants sont, comme démontré avec les essais expérimentaux, que l'albumine, la protéine étudiée, empêche les dégradations de l'acier inoxydable aux plus faibles concentrations d'ions chlorure ; ensuite, aux plus fortes concentrations de chlorures, un temps d'incubation est nécessaire pour détruire le film passif.

## Introduction

Le nombre de prosthèses totales de hanche augmentera dans les prochaines années. En effet, il a doublé entre 2000 et 2010, en France [1]. Aux Etats-Unis, 250000 prothèses totales de hanche sont implantées annuellement, en 2012. Il est envisagé que ce nombre passe à 572000 en 2030. Il est important de noter le nombre de prosthèses totales de hanche devrait aussi augmenter d'un facteur 2 en 2030. De plus 1 américain sur 30 possède une prothèse totale de hanche. Enfin, les politiques de santé sur la mobilité concernent des patients de plus en plus jeunes. [2]. Tous ces éléments vont dans le sens d'une augmentation drastique du nombre d'implants orthopédiques. Un des points clef lié à la prévision de la durée de vie des implants orthopédiques est du ressort de la compréhension de l'usure pour prédire la stabilité de l'implant et éviter la production de débris qui est responsable du rejet de l'implant. Quand une prothèse est implantée pour la première fois, la première étape liée à l'ostéointégration consiste en l'adhérence entre les tissus osseux et le matériau étranger. Après cette étape, en raison du stress shielding (masquage des contraintes), un descellement intervient et des micro-mouvements se produisent entre la tige fémorale et l'os. Ce dernier point implique des micro-mouvements dans un environnement aqueux corrosif, ce qui est appelé fretting corrosion [3]. Comme indiqué précédemment, même si le métal a de meilleures performances mécaniques que celles de l'os, en fretting corrosion l'usure de l'acier inoxydable est plus grande que celle du polymère, pourtant beaucoup moins dur que l’acier inoxydable. Avant d’entreprendre des tests très proches des conditions réelles de sollicitation des prostheses totals de hanche, avec notamment la prise en compte du milieu physiologique complexe, des tests entre l’acier inoxydable 316L et le Polyméthacrylate de méthyle (PMMA) ont été entrepris ; ces matériaux ont des caractéristiques mécaniques très proches de celles des implants orthopédiques et de l’os [4,5]. Plusieurs résultats intéressants ont été fournis par des tests expérimentaux et les auteurs impliqués dans ce travail souhaitent apporter de nouvelles connaissances en utilisant une modélisation basée sur le Point Defect Model [6-8]. L’innovation consiste à ajouter une contribution du fretting au PDM, modèle prédictif et déterministe. C’est la première fois, comme les tests de fretting corrosion avec un contrôle des paramètres électrochimiques et mécaniques, que ce type d’approche est proposé. Plusieurs expériences de fretting corrosion à potentiel libre et à potentiel imposé ont été entreprises La démarche a été appliquée à ces résultats expérimentaux pour obtenir des résultats grâce au PDM amélioré de la contribution en fretting.

Ce travail vise à investiguer les résultats provenant des expériences de spectroscopie d’impédance électrochimiques à différents temps d’une même expérience et ceux modélisés pour obtenir les multiples constantes physiques du problème. Alors, l’épaisseur du film passif a été calculée. La methode de calcul comporte de nombreuses approximations. La première consiste à considérer que la partie anodique, à potentiel libre de corrosion, est due à la zone d’usure et que la partie cathodique est la zone extérieure à cette zone d’usure. Cette hypothèse est fondée sur le principe que le courant est principalement dû à la zone d’usure soumise au fretting corrosion.

L’épaisseur du film passif attendue doit être considérée comme un indicateur pour savoir si la couche passive existe ou non en présence d’albumine, film fin à la surface du film passif. Connaître avec la précision du nanomètre l’épaisseur du film passif ne serait pas raisonnable. Cependant cette approche est une bonne indication pour l’influence des conditions expérimentales sur la passivité de l’acier inoxydable étudié.

# Résultats

L’approche expérimentale du fretting corrosion est décrite dans [4,5]. De ces investigations des résultats ont été obtenus sur l’influence de la concentration de chlorure avec une certaine teneur de protéines, i.e. l’albumine dans ce cas. Le contact étudié est l’acier inoxydable 316L, de forme parallélépipédique, contre un échantillon de PMMA, forme cylindrique. Le protocole de polissage a une grande importance, ainsi il est décrit précisément dans [9-10]. Les solutions sont composées de solutions de NaCl, d’une concentration allant de $10^{-3}$ mol.L$^{-1}$, $10^{-2}$ mol.L$^{-1}$, $10^{-1}$ mol.L$^{-1}$ à 1 mol.L$^{-1}$. 2 types de solutions salines ont été envisagées dans cette étude : sans albumine et avec, concentration de 20 g.L$^{-1}$. La modélisation tiendra compte de ces résultats expérimentaux.

Les résultats expérimentaux pertinents sont les diagrammes de Nyquist provenant des expériences de spectroscopie d’impédance électrochimique. Durant les tests de fretting corrosion, toutes les 20 minutes, les diagrammes de Nyquist ont été enregistrés. La Figure 1 illustre les différents diagrammes obtenus lors d’une experience de fretting corrosion à potential imposé, -400mV/ECS, 20 g.L$^{-1}$ d’albumine et 1 M de NaCl. Les diagrammes de Nyquist évoluent dans le sens d’une augmentation de la partie imaginaire, comme indiqué sur la Figure 1. La partie imaginaire, en valeur absolue, augmente donc avec le temps lors d’une expérience, ce qui peut être interprété par l’effet protecteur de l’albumine qui s’adsorbe au cours de l’expérience. A partir de ces résultats il est possible de comparer les résultats expérimentaux et ceux provenant de la modélisation grâce au PDM adapté au cas du fretting corrosion

Dans cette section la procédure de modélisation va être décrite à partir du PDM. La Figure 2 résume les sept réactions du PDM qui seront utilisées dans le cadre de cette étude [11]. La dissolution du metal est prise en compte dans la partie anodique du circuit électrique équivalent. La relation (1) décrit l’épaisseur du film stationnaire dans des conditions de pseudo stationnarité. Seules les réactions 3 and 7 sont non conservatives et elles sont impliquées dans l’expression de l’épaisseur de film d’oxydes dans des conditions stationnaires. Le terme W est la contribution du fretting, usure mécanique. Ce terme est issu des mesures expérimentales et donné par la loi d’Archard [12].

$$\frac{\mathrm{d}L}{\mathrm{d}t} = \Omega k_3^0\, \mathrm{e}^{a_3 V}\, \mathrm{e}^{b_3 L_{ss}}\, \mathrm{e}^{c_3 pH} - \Omega k_7^0\, \mathrm{e}^{a_7 V}\, \mathrm{e}^{c_7 pH} \left(\frac{C_{\mathrm{H}^+}}{C_{\mathrm{H}^+}^0}\right)^n - W = \frac{\mathrm{d}L^+}{\mathrm{d}t} - \frac{\mathrm{d}L^-}{\mathrm{d}t} - W \quad [1]$$

Le coefficient de la loi d’Archard est un paramètre qui sera ajusté lors de la modélisation.

Pour chaque spectre EIS, le potentiel est fixé et le courant (ou la densité de courant) est mesurée. De ce point le PDM peut être utilisé dans l’expression (2):

$$I = F \left[ \begin{array}{l} \chi k_1^0\, \mathrm{e}^{a_1 V}\, \mathrm{e}^{b_1 L_{ss}}\, \mathrm{e}^{c_1 pH}\, C_{V_M}^{m/bl} + \chi k_2^0\, \mathrm{e}^{a_2 V}\, \mathrm{e}^{b_2 L_{ss}}\, \mathrm{e}^{c_2 pH} + \chi k_3^0\, \mathrm{e}^{a_3 V}\, \mathrm{e}^{b_3 L_{ss}}\, \mathrm{e}^{c_3 pH} + \\ (\delta - \chi) k_4^0\, \mathrm{e}^{a_4 V}\, \mathrm{e}^{c_4 pH} + (\delta - \chi) k_5^0\, \mathrm{e}^{a_5 V}\, \mathrm{e}^{c_5 pH}\, C_{M_i}^{bl/ol} + \\ (\delta - \chi) k_7^0\, \mathrm{e}^{a_7 V}\, \mathrm{e}^{c_7 pH} \left(\frac{C_{\mathrm{H}^+}}{C_{\mathrm{H}^+}^0}\right)^n \end{array} \right] \quad [2]$$

The potential is imposed and the current density is measured, thus one may suggest that an equivalent electrical circuit could be suggested.

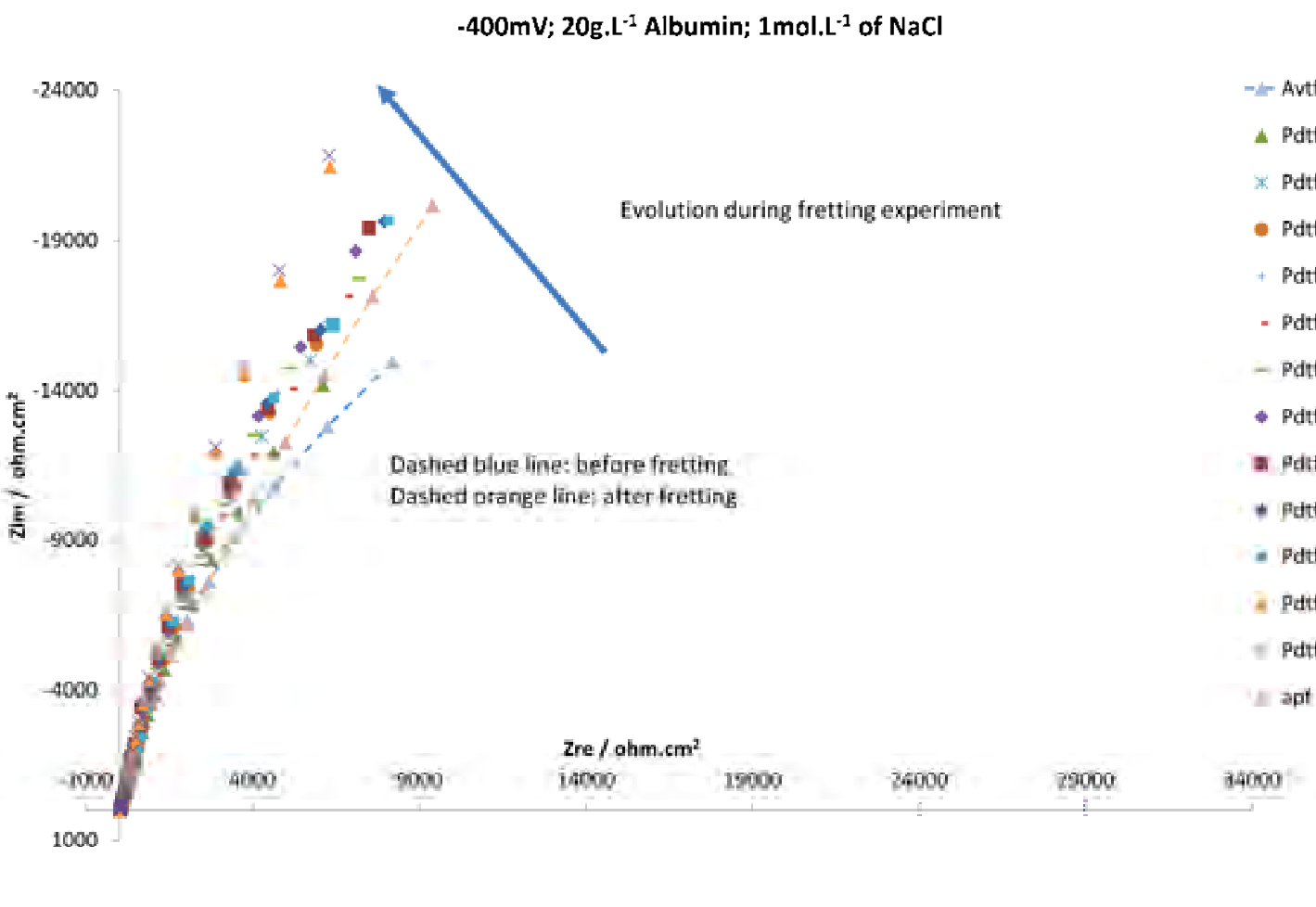

1.E-07

0.E+00

*Figure 1: Diagrammes de Nyquist obtenus lors d'une experience de fretting corrosion, potentiel imposé -400 mV/ECS, 1 mol.L$^{-1}$ de NaCl et 20 g.L$^{-1}$ d'albumine. Avtf: avant fretting; Apf: après fretting, Pdtf(i): pendant l'expérience de fretting corrosion*

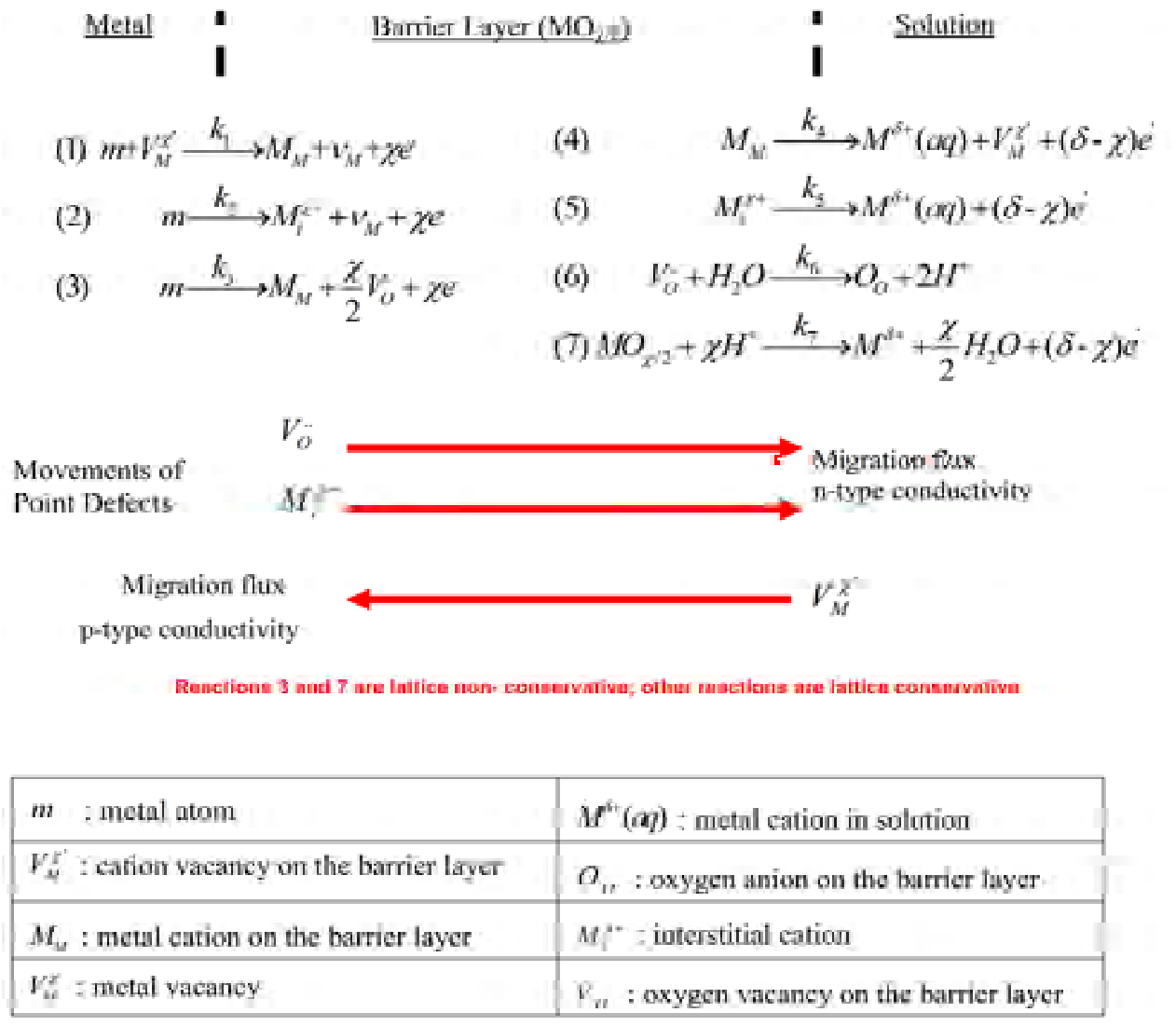

*Figure 2 7 réactions du Point Defect Model, approche déterministe [7].*

Le circuit électrique consiste en une partie anodique (une impédance liée à la dissolution du métal et d'une impédance de Warburg) et une partie cathodique (association en parallèle de la réduction du dioxygène dissous et de celle du protons) et une élément à phase constant correspondant à la double couche à la surface du métal. L'innovation, en termes de modélisation, consiste en l'utilisation de l'algorithme génétique pour trouver les 29 paramètres du PDM. La méthode classique, comme celle de Newton-Raphson, ne permet pas de trouver facilement un minimum global dans un espace de solutions aussi grand. Les avantages de l'algorithme génétique résident dans le fait de sélectionner les solutions optimales à chaque itération.

Au potentiel libre de corrosion, l'épaisseur de film passif en régime quasi stationnaire, Lss, pourra être calculée à partir du modèle. Une des preuves expérimentales de la disparition du film passif est le dégagement gazeux observé sur les enregistrements vidéo effectués pendant les tests de fretting corrosion. En effet, le dégagement gazeux est la preuve d'une acidification de la surface de contact, comme prédit dans [10]. La Figure 4 met en évidence deux tendances à partir des calculs de Lss issus de l'algorithme génétique. La Figure 3 a) [11] met en évidence l'évolution des oxydes selon le temps durant une expérience. Il est important de noter qu'un temps d'incubation est nécessaire pour pour pouvoir calculer par cette modélisation la diminution du film passif. Au-delà de 100 minutes, le calcul nous montre que la couche d'oxydes diminue, Figure 3 a), 10$^{-2}$ mol.L$^{-1}$ de NaCl et 0 g.L$^{-1}$ d'albumine. Figure 3 b), la concentration de NaCl est la même concentration que celle de la Figure 3 a) mais la comcentration d'albumine est 20 g.L$^{-1}$. Sur la Figure 3, Lss est plus grande en présence d'albumine que sans. Il est à noter que le temps d'incubation semble être le même, avec et sans albumine.is a little bit higher than the one without albumin. Les épaisseurs modélisées et les faits expérimentaux Ainsi les épaisseurs de films d'oxydes et les faits expérimentaux tendent à montrer que l'albumine protège l'acier inoxydable 316L contre les dégradations par fretting corrosion. Les faits expérimentaux ont été mis en évidence dans [5].

a)

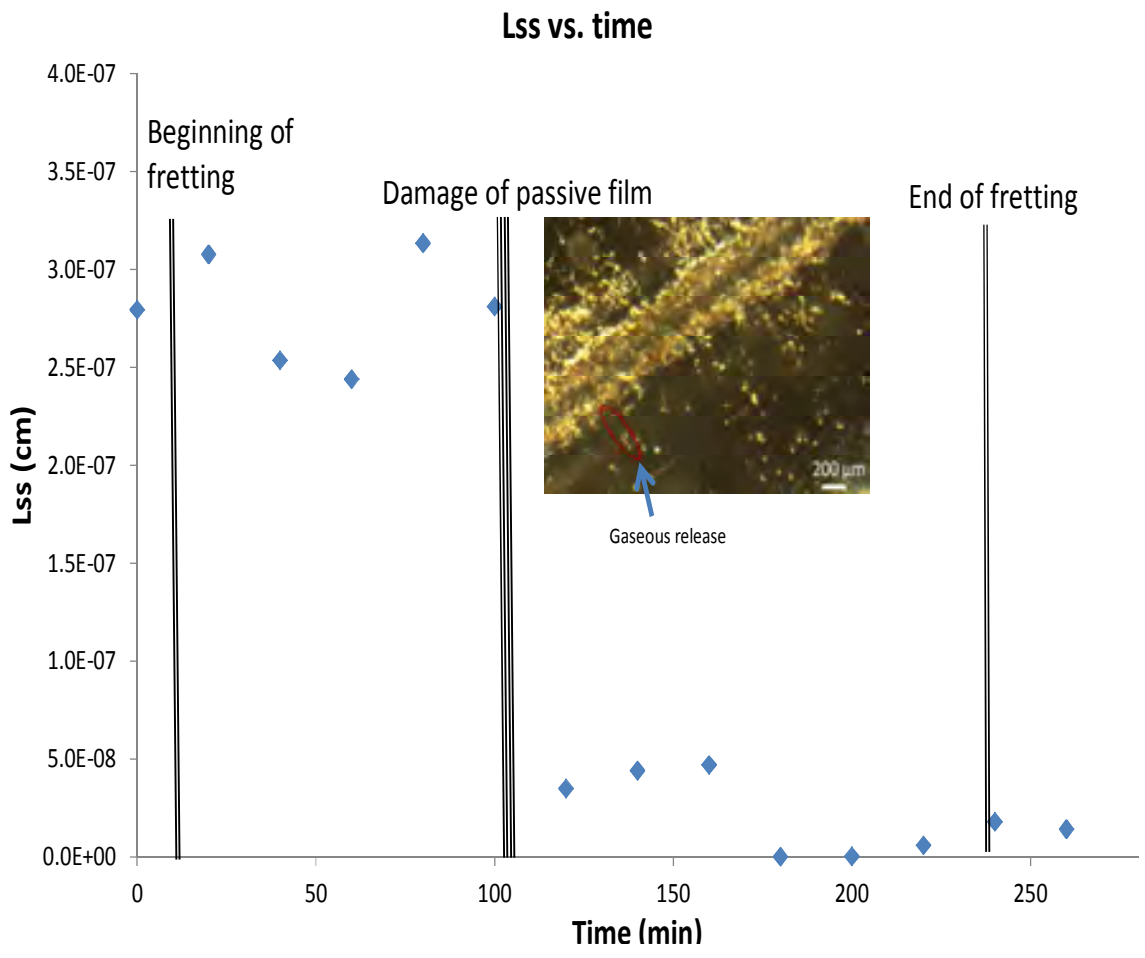

b)

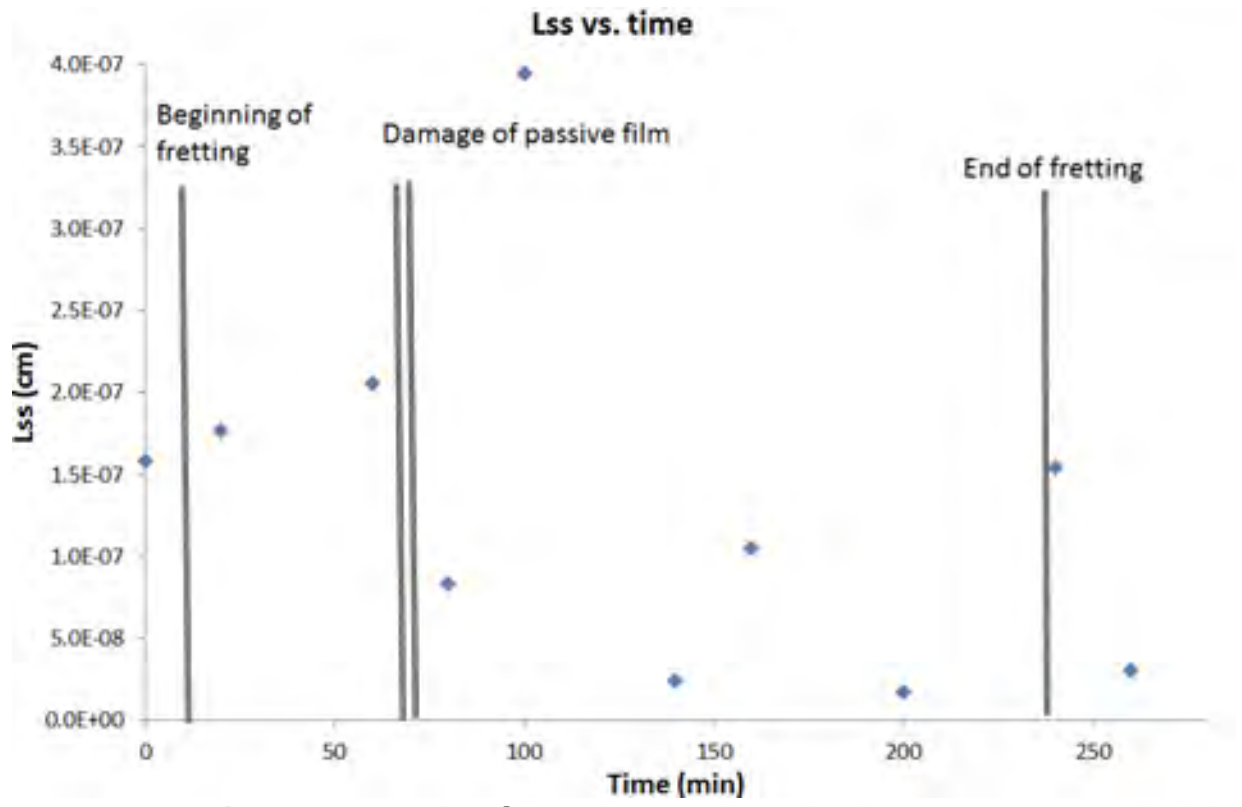

*Figure 3 épaisseur du film d'oxydes de l'acier inoxydable 316L en fonction du temps durant une expérience de fretting corrosion, 10$^{-2}$ mol.L$^{-1}$, 14,400 secondes a) modélisation sans albumine b) avec albumine.*

La Figure 4 rassemble les résultats obtenus des épaisseurs de film d'oxydes calculés grâce au PDM en tenant en compte le fretting. L'incertitude, i.e. l'intervalle de confiance à 95%, des résultats issus de la modélisation restent grands. Il est d'environ 2 nm. Comme indiqué précédemment, le temps d'incubation est d'environ 100 minutes. Cet élément doit être pris en compte pour expliquer la grande incertitude de ces résultats. C'est la raison pour laquelle la figure 5 représente les épaisseurs calculées après un temps de fretting corrosion de 120 minutes. Tout d'abord à 10$^{-3}$ mol.L$^{-1}$, l'albumine a un effet protecteur de la couche d'oxydes. Ce résultat est en accord avec la mesure du volume d'usure le plus faible en présence d'albumine faible a 0,1 mol.L$^{-1}$ de force ionique, la tendance semble inverse : l'albumine ne protège pas le métal. Il est à noter que cette concentration est seuil. De nombreux phénomènes physiques doivent se produire à ce seuil, ainsi le modèle choisi pourrait être mis en défaut. Le résultat le plus intéressant concerne la solution de Ringer. L'albumine protège le 316L durant une dégradation par fretting corrosion.

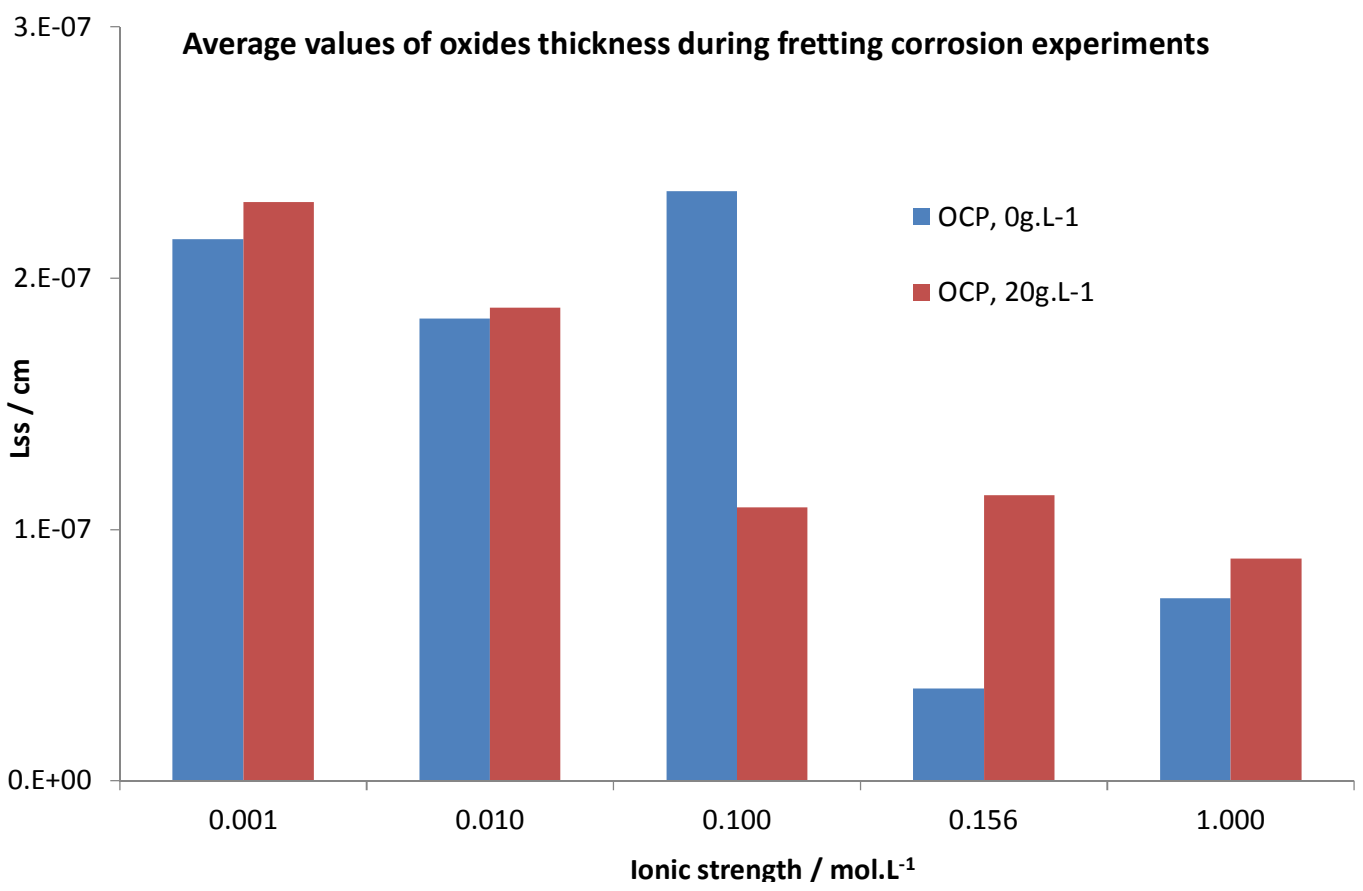

*Figure 4: Valeurs moyennes de l'épaisseur de la couche d'oxydes calculées grâce au PDM-fretting. 0,156 M de force ionique pour la solution de Ringer (NaCl : 8,5 g.L$^{-1}$, KCl : 0,25 g.L$^{-1}$, $CaCl_2$, $2H_2O$: 0,22 g.L$^{-1}$, $NaHCO_3$ : 0,15 g.L$^{-1}$); ces valeurs proviennent des 12 SIE enregistrés durant 1 expérience de fretting corrosion.*

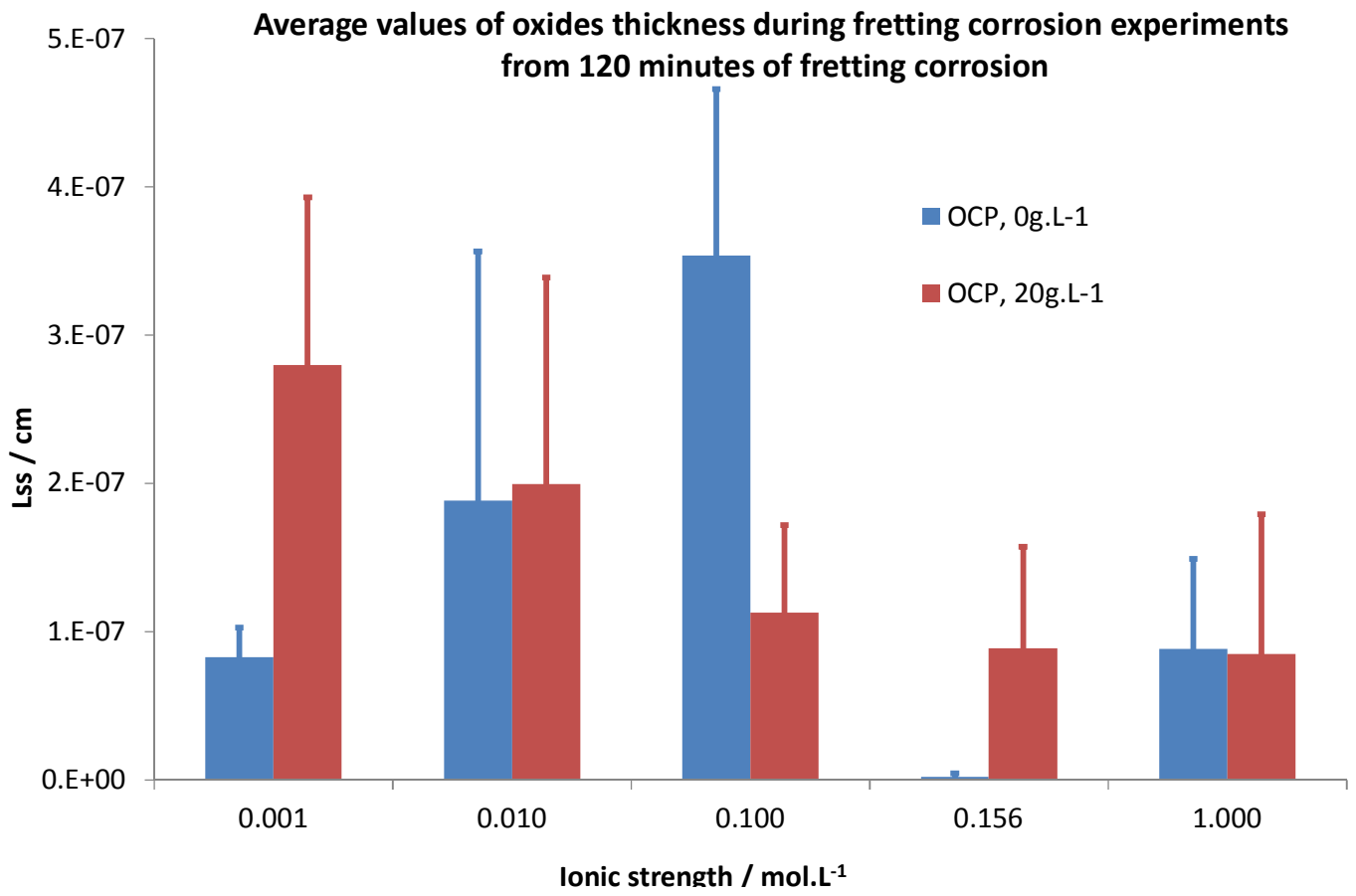

*Figure 5: Les valeurs moyennes de la couche d'oxydes calculées par le modèle PDM-fretting. 0.156 mol.L$^{-1}$ de force ionique correspond à celle de la solution de Ringer* (NaCl : 8,5 g.L$^{-1}$, KCl : 0,25 g.L$^{-1}$, $CaCl_2$, $2H_2O$: 0,22 g.L$^{-1}$, $NaHCO_3$ : 0,15 g.L$^{-1}$); *ces valeurs sont issues des 7 dernières expériences d'EIS, après un temps d'incubation de 100 minutes.*

# Conclusions:

L'épaisseur quasi stationnaire du film passif a été calculée grâce au PDM-fretting. Deux points importants ont été mis en évidence dans cette étude. En premier lieu, un temps d'incubation a été mis mis en lumière pour tous les tests de fretting corrosion.Le second point concerne l'effet protecteur de l'albumine sur la dégradation de l'acier inoxydable 316L. Pour une concentration de 1 mol.L$^{-1}$ de NaCl, l'albumine n'a pas d'effet sur l'épaisseur du film d'oxydes. Un effet protecteur a été mis en évidence par cette modélisation pour la plus faible concentration de NaCl, i.e. $10^{-3}$ mol.L$^{-1}$. Cette modélisation a été entreprise pour la première fois, dans cette étude. La modélisation grâce aux algorithmes génétiques a permis d'obtenir des résultats dans des délais raisonnables de modélisation, i.e. moins d'une journée sur un ordinateur de bureau amélioré.

Plusieurs investigations sont en cours pour affiner la modélisation par les algorithmes génétiques. En termes de matériaux, une nouvelle campagne est en cours pour étudier le comportement de l'alliage de Co-Cr-Mo en fretting corrosion.

# Références